\documentclass[prb,twocolumn,showpacs]{revtex4}
\usepackage[dvips]{graphicx}

\begin{document}
\title{Aluminum Oxide Layers as Possible Components for Layered Tunnel Barriers}
\author{E. Cimpoiasu$^{1}$, S. K. Tolpygo$^{1}$, X. Liu$^{1}$, N. Simonian$^{1}$, J. E. Lukens$^{1}$, R. F. Klie$^{2}$, Y. Zhu$^{2}$, and K. K. Likharev$^{1}$}
\affiliation{$^{1}$Stony Brook University, Stony Brook, NY 11794-3800\\
$^{2}$Brookhaven National Laboratory, Upton, NY 11973-5000}
\date{\today}

\begin{abstract}

We have studied transport properties of Nb/Al/AlO$_{x}$/Nb tunnel junctions with ultrathin aluminum oxide layers formed by (i) thermal oxidation and (ii) plasma oxidation, before and after rapid thermal post-annealing of the completed structures at temperatures up to 550$^\circ$C.  Post-annealing at temperatures above 300$^\circ$C results in a significant decrease of the tunneling conductance of thermally-grown barriers, while plasma-grown barriers start to change only at annealing temperatures above 450$^\circ$C.  Fitting the experimental I-V curves of the junctions using the results of the microscopic theory of direct tunneling shows that the annealing of thermally-grown oxides at temperatures above 300$^\circ$C results in a substantial increase of their average tunnel barriers height, from $\sim$1.8 eV to $\sim$2.45 eV, versus the practically unchanged height of $\sim$2.0 eV for the plasma-grown layers.  This difference, together with high endurance of annealed barriers under electric stress (breakdown field above 10 MV/cm) may enable all-AlO$_{x}$ and SiO$_{2}$/AlO$_{x}$ layered "crested" barriers for advanced floating-gate memory applications.
\end{abstract}

\maketitle

Calculations\cite{1,2,3} indicate that tunnel conductance of layered barriers, in particular those with "crested" potential profile peaking in the middle, may be much more sensitive to the applied voltage than that of the uniform layers.\cite{4}  This sensitivity, if combined with high endurance to electric stress, may be used in advanced floating-gate memories, including fast and scalable random access memories \cite{5} and fast single- and few-electron memories,\cite{6,7} and ultradense data storage systems,\cite{1} as well as for improvement of the usual nonvolatile (e.g., "flash") memories.\cite{8,9} However, finding an appropriate combination of materials for crested barrier layers presents a challenge.  Indeed, numerous experiments (for a review see, e.g., Ref.[10]) indicate that just a few known CMOS-compatible materials may combine the barrier height sufficient for thermionic current suppression at room temperature (above $\sim$1.5 eV), with the necessary high breakdown field (above 10 MV/cm), and negligible trap-assisted tunneling.  To our knowledge (see also the recent theoretical calculations)\cite{11}, the list of such candidate materials is essentially limited to: (i) silicon dioxide, (ii) low-trap-density silicon nitride that may be grown using special methods,\cite{12,13} and (iii) aluminum oxides grown by a variety of methods including notably thermal \cite{14} and plasma \cite{15} oxidation.

The goal of this paper is to show that the aluminum oxides represent a good material choice for fabrication of crested barriers.  Experimental measurements of the most important parameter in this context, the average tunnel barrier height $\langle U \rangle$, have been reported for aluminum oxide layers in quite a few publications.  Unfortunately, the cited values of $\langle U \rangle$,  are scattered rather broadly: for thermally-grown oxides, most results are in the range from 1.7 to 2.5 eV (see, e.g., Refs. [16-22]), but values as low as 1.2 eV,\cite{23} and as high as 4.75 eV,\cite{24} or even 20 eV,\cite{25} have also been derived from the data.  Similarly, for plasma-grown layers, most reported values of $\langle U \rangle$,  are in the range from 1.7 to 2.3 eV (see, e.g., Refs. [24], [26-30]),\cite{31} but numbers as high as 3.6 eV have also been claimed.\cite{29}  The published results for the apparent barrier asymmetry, $\Delta U \equiv U_{max}-U_{min}$, are scattered even more, from a few tenths of eV all the way up to 6 eV,\cite{24} and the only apparent consensus is that the barrier is always higher at the top (counter-electrode) interface.  Probably, the most important source of these differences are those of the film fabrication, including the substrate temperature (that has not always been carefully monitored) and the counter-electrode material.  However, some result scattering may be also attributed to the variety of techniques used for barrier height measurement, including $I$-$V$ curve fitting,\cite{16,17,19,18,20,21,22,24,25,27,28,29,30} photoelectric effect,\cite{18,20,26} and ballistic electron emission spectroscopy.\cite{23}  Some of these methods may give rather inaccurate results.  For example, as it has been shown in our recent work,\cite{22} fitting of $I$-$V$ curves of aluminum oxide barriers with WKB approximation results may lead to substantial errors, since such barriers are rather thin and sharp.  These errors may be dramatically increased if low-$V$ expansions of WKB formulas\cite{32,33} are used, as this procedure is highly vulnerable to minor additional currents due to inelastic (e.g., trap-assisted) tunneling - see, e.g., Fig. 2 and its discussion below.  Our experience shows that fitting the slope of the "Fowler-Nordheim plot" ($lnI$ vs. $1/V$) of high-$V$ data may also lead to very substantial errors.  The experimental information on the effective mass of the tunneling electrons is even more limited (see the discussion below).

One more motivation for additional experimentation was to study the effects of rapid thermal post-annealing of tunnel junctions.  It was noticed previously that thermal annealing may improve tunneling magnetoresistance of junctions between magnetic layers\cite{34,35,36,37,38,39,40} and, for high annealing temperatures $T_a$, change the atomic structure of the oxide quite substantially\cite{41}.  However, the annealing effect on the barrier height $\langle U \rangle$,  has not been studied in any detail, to the best of our knowledge. (Some changes in $\langle U \rangle$,  at $T_a<300^{\circ}$C were noticed in Refs. [34], [36], and [38-40], but the uncertainty of the results, obtained using the WKB expansion,\cite{32,33} was comparable with the change itself.)  Thus, we have carried out detailed studies of tunnel barriers grown by thermal and plasma oxidation of aluminum, and rapid post-annealed at various temperatures. 

\begin{figure}[tbh]
\includegraphics[width=6.5cm,angle=270]{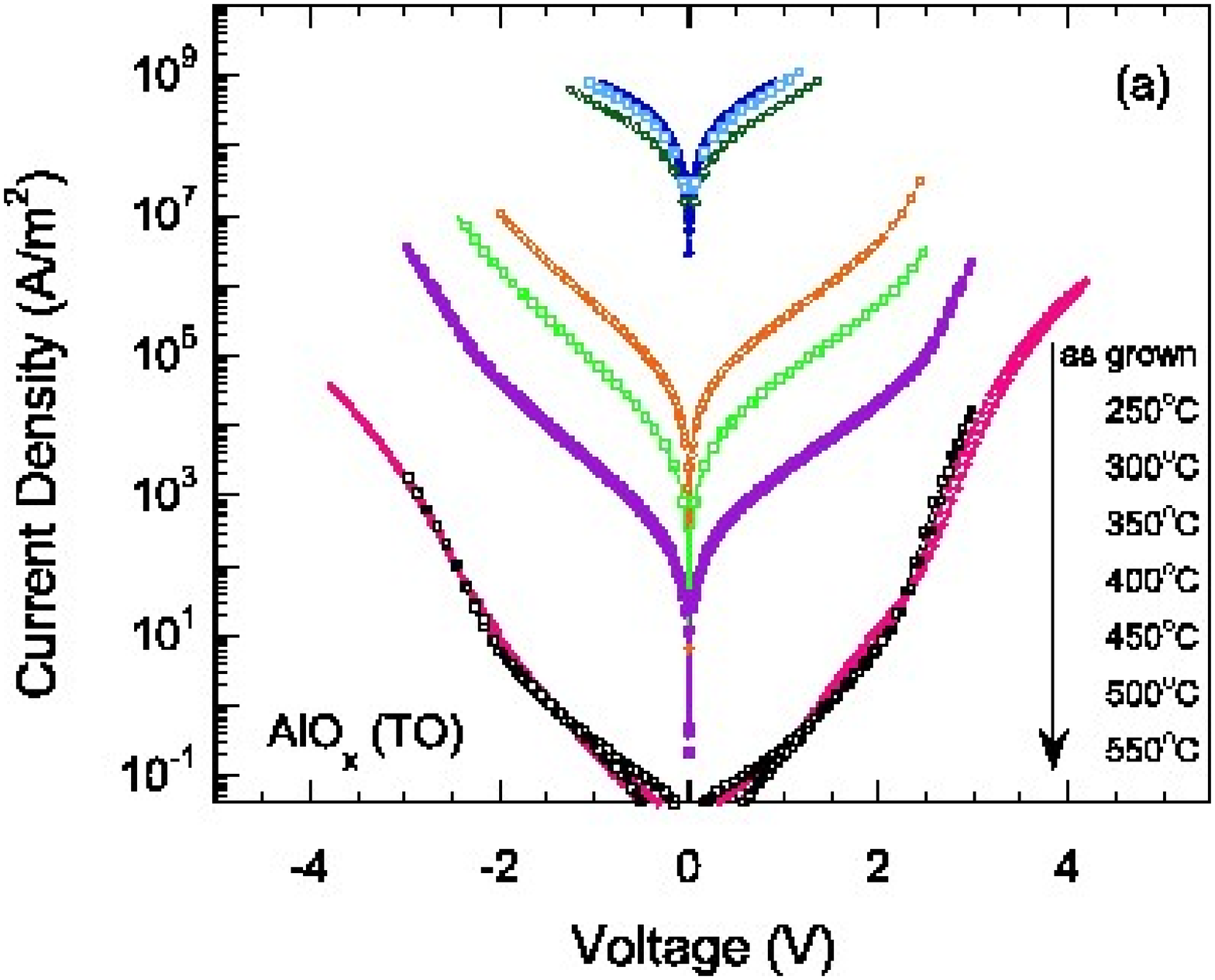}
\includegraphics[width=6.5cm,angle=270]{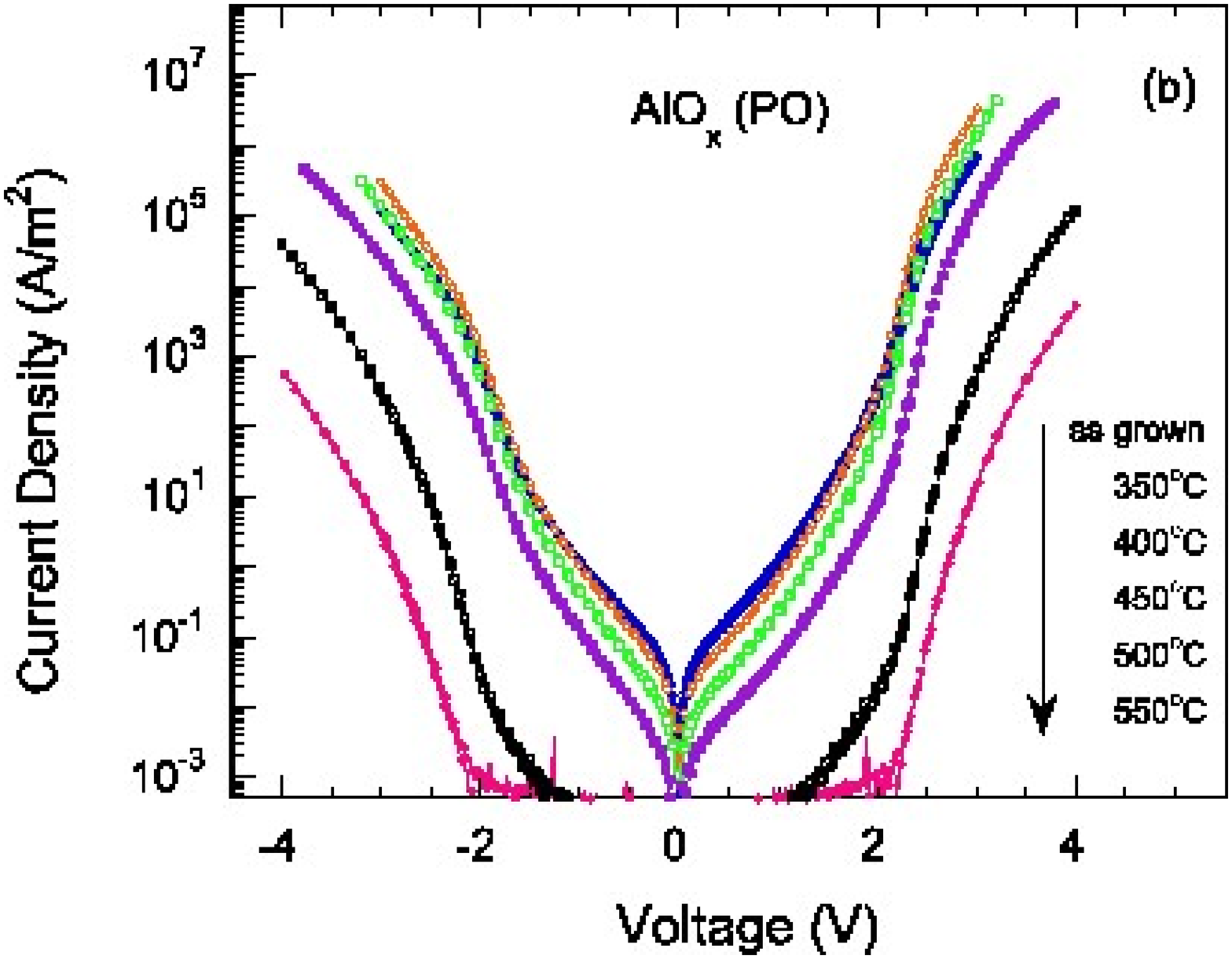}
\caption{Experimental current density $J$ as a function of the applied dc voltage $V$ for Nb/Al/AlO$_x$/Nb junctions with (a) thermally-grown (Crest 5) and (b) plasma-grown (Crest 19) oxide layers, before and after the rapid thermal anneals at indicated temperatures, as measured at 4.2 K.  The noisy flattening of the lower curve in panel (b) at small voltages is due to leakage of our measurement setup at I $\le 10^{-13}$ A. }
\end{figure}

The oxide layers have been grown on oxidized Si wafers ($\rho \approx 10\:\Omega$-cm) covered by 500 nm of thermally-grown SiO$_2$, as components of standard Nb-trilayer junctions.\cite{42}  The oxides were formed in-situ on 10-nm-thick aluminum films that had been dc-magnetron-sputtered on similarly deposited 150-nm-thick niobium films, using either exposure to dry oxygen or in 13.6 MHz oxygen plasma, both at room temperature.  (Wafers were kept on a water-cooled, dc-insulated holder.)  After in-situ sputtering of a niobium 100-nm-thick counter-electrode and sample patterning into junctions of various area $A$ (3$\times$3, 30$\times$30, and 300$\times$300 mm$^2$), a few chips from each wafer were subjected to rapid thermal annealing.  $DC$ $I$-$V$ measurements of both as-oxidized and annealed junctions have been carried out at both room and helium (4.2 K) temperatures, using a special low-noise, high-sensitive setup.  Voltage sweeps with gradually growing amplitude were used to characterize transport up to the very onset of hard breakdown. 

Here we focus on comparing the results from two representative wafers: "Crest 5" (thermal oxidation for 40 minutes at 100 Torr) and "Crest 19" (plasma oxidation for 3 minutes at 15 mTorr), both post-annealed at temperatures up to 550$^{\circ}$C in inert atmosphere (either Ar or N$_2$).\cite{43}  The junctions, both before and after post-annealing, were highly reproducible, with the r.m.s. on-chip (junction-to-junction) variation of low-voltage conductance from as low as 0.8\% (considerably better than any published results we are aware of ) to ~20\% (comparable with the reported results - see, e.g., Ref. [44]).

\begin{figure}[tbh]
\includegraphics[width=6.5cm,angle=270]{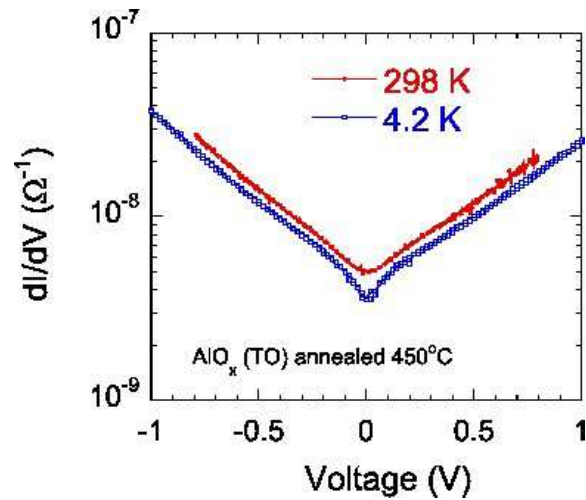}
\caption{The nonlinear "dynamic" conductance $G(V) \equiv dI/dV$ of a typical 450$^{\circ}$C-annealed sample from wafer Crest 5, measured at liquid-helium and room temperatures.  }
\end{figure}

Figure 1 shows $I$-$V$ curves of representative junctions from these two wafers, both before and after post-annealing at various temperatures.  (These data have been taken at helium temperature; however, the increase of temperature to 300 K changes the current only slightly - see Fig. 2).  First of all, one can see that the annealing above $\sim$300$^{\circ}$C leads to a considerable improvement of the junction quality: the hard breakdown voltage V$_b$ increases, and the $I$-$V$ curves show virtually no hysteresis or "soft breakdown", up to V$_b$. (For thermally-grown oxides annealed at $\sim$500$^{\circ}$C and beyond, the hysteresis appears again, though V$_b$ continues to grow.)  More quantitatively, the charge to breakdown, measured at room temperature for samples annealed at 450$^{\circ}$C, stays above 10$^5$ C/cm$^2$ (i.e., a few orders of magnitude higher than the level typical for industrial grade SiO$_2$ barriers)\cite{14} until $\sim$2.1 V for both Crest 5 and Crest 19 samples.

However, our most important observation is a dramatic difference between the effects of annealing temperature upon the thermally-grown and plasma-grown oxides: while the low-voltage conductance of the former junctions drops sharply starting above $\sim$300$^{\circ}$C and reaches almost 6 orders of magnitude by 450$^{\circ}$C,\cite{45} the reduction in the latter barriers is minor (below 2 orders of magnitude) until the annealing temperature has been raised to $\sim$500$^{\circ}$C.  In order to give at least a phenomenological interpretation of this effect, we have used theoretical fits to extract essential tunnel barrier parameters of the AlO$_x$ layers.

\begin{figure}[tbh]
\includegraphics[width=6.5cm]{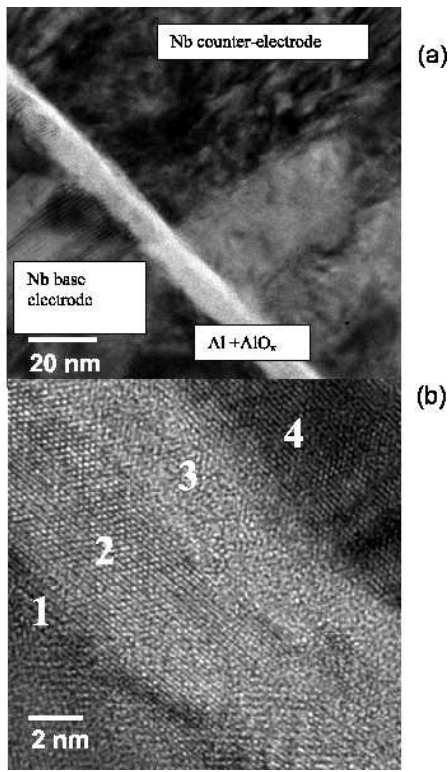}
\caption{(a) High-resolution transmission-electron-microscope images of a 450$^{\circ}$C post-annealed sample from wafer Crest 5 for two different magnifications and (b) magnified part of the layered structure with the position of the electron energy loss spectroscopy spectra indicated. }
\end{figure}

\begin{figure}[tbh]
\includegraphics[width=6.5cm]{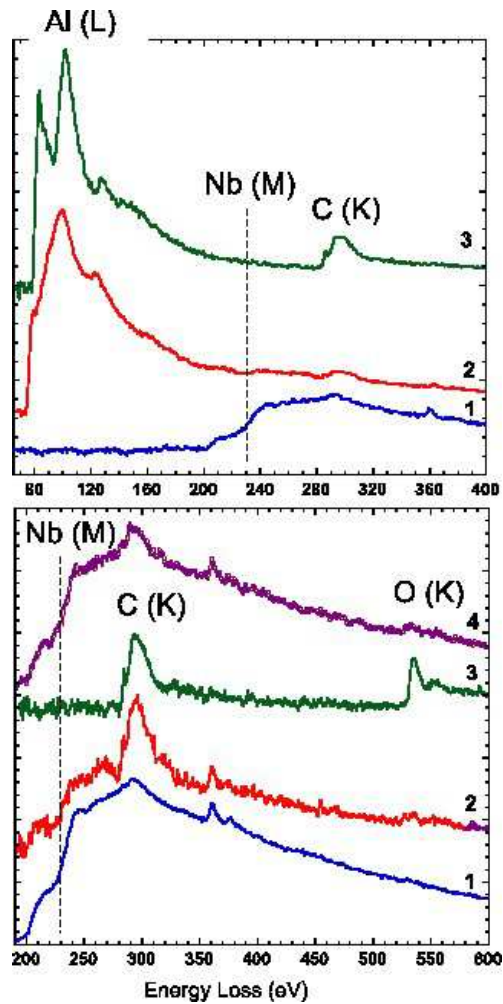}
\caption{Results of the electron energy loss spectroscopy for two energy ranges from the positions indicated in Fig. 3b: 1 - Nb base electrode, 2 - the middle of Al layer, 3 - AlO$_x$ layer, and 4 - Nb counter-electrode.  The spectra are background subtracted, and corrected for multiple scattering contributions. The carbon K-edge present in all the spectra stems from the carbon build-up during the spectrum acquisition and is not a feature of the sample structure. }
\end{figure}

The shape of $I$-$V$ curves of the samples (Fig. 1), and their very weak temperature dependence (Fig. 2) are consistent with the assumption of direct tunneling of electrons through the barrier.\cite{46}  [This conclusion is also supported by the results of high-resolution microscopy (Fig. 3) and electron energy loss spectroscopy (Fig. 4) of the annealed samples, showing a well-defined oxide layer with sharp interfaces with both base and counter-electrodes.]  This is why we have fitted our experimental data with results of a "microscopic" (non-WKB) theory of such tunneling.  Our general computer algorithm is based on the joint solution of the 1D Schr\"odinger equation (using the transfer-matrix technique) and Poisson equation for tunneling electrons.\cite{47}  However, we have found that the barrier shape modification by the charge of tunneling electrons is very small.  The exclusion of the Poisson solver from the code makes it very fast: simulation of one $I$-$V$ curve in $\sim$100 points with a few-percent accuracy takes about 1 minute on a single-processor workstation.  The code has been checked on the results for SiO$_2$ layers described in Ref. [48], and gave similar results to those of the theoretical calculations in that seminal paper. 

Figure 5 shows the results of the fitting of the voltage dependence of the specific dynamic conductance $g(V) \equiv A^{-1}dI(V)/dV$ for junctions of both types post-annealed at 450$^{\circ}$C.  The advantage of fitting the semilog $g(V)$ plots rather than the $lnI(V)$ curves (Fig. 1) is that in the former case the peculiarities of low-voltage behavior are revealed more clearly - see also Fig. 2.  They show, in particular, a minor "cusp" contribution $G_h\propto |V|^{\alpha-1}$ to the conductance (and hence $I_h \propto sgn(V)|V|^{\alpha}$ to current,) similar to that observed and discussed by others - see, e.g., Refs. [46], [49-51].  Figure 2 shows that this current component is more sensitive to temperature than the current at higher voltages, though this temperature dependence is still much weaker than that for the Poole-Frenkel conductance mechanism.\cite{52} Though the exact identification of the physics of the $I_h$ component is beyond the scope of this work, we believe that it is due to some sort of hopping (trap-assisted tunneling) strongly affected by the Coulomb interaction of the hopping electrons.  In fact, it may be best fitted with the values ($\alpha$ = 1.8$\pm$0.1 for Crest 5 and $\alpha$ = 2.0$\pm$0.3 for Crest 19) that are relatively close to that of the classical Mott-Gurney law ($\alpha$ = 2)  for space-charge-limited current.\cite{53}  A better agreement would be hard to expect, since the Mott-Gurney model implies that the layer thickness $d$ is much larger than the localization radius $a$ of a typical trap, and the thickness of our barriers ($d \sim$ 2 to 3 nm as shown below) is comparable with  the estimated value $a\sim$ 1 nm.  For the best fitting, we have subtracted $I_h$ from the data (see the dashed lines in Fig. 5), although the fitting results are appropriate even for the raw data.

As Fig. 5 shows, a relatively good fitting of the data may be achieved with the traditional trapezoidal (i.e., one-layer) model of the barrier.  However, better fitting is provided by the potential profile approximation with two (for Crest 5) or three (for Crest 19) linear pieces, implying a layered structure of the oxide.  This is not too surprising, since the complex interface chemistry, as well as trapped charge impurities (see, e.g., Ref. [54]) may provide interfacial layers with properties different from the oxide bulk.  Note that while the $I$-$V$ curve fitting gives very definite results for the effective thickness $d_{ef} = (m/m_0)^{1/2}d$ of the layers, it cannot distinguish the contributions to $d_{ef}$ from the effective mass $m$ of the tunneling electron and from the physical thickness $d$ of the barrier.\cite{55} 

\begin{figure}[tbh]
\includegraphics[width=6.5cm,angle=270]{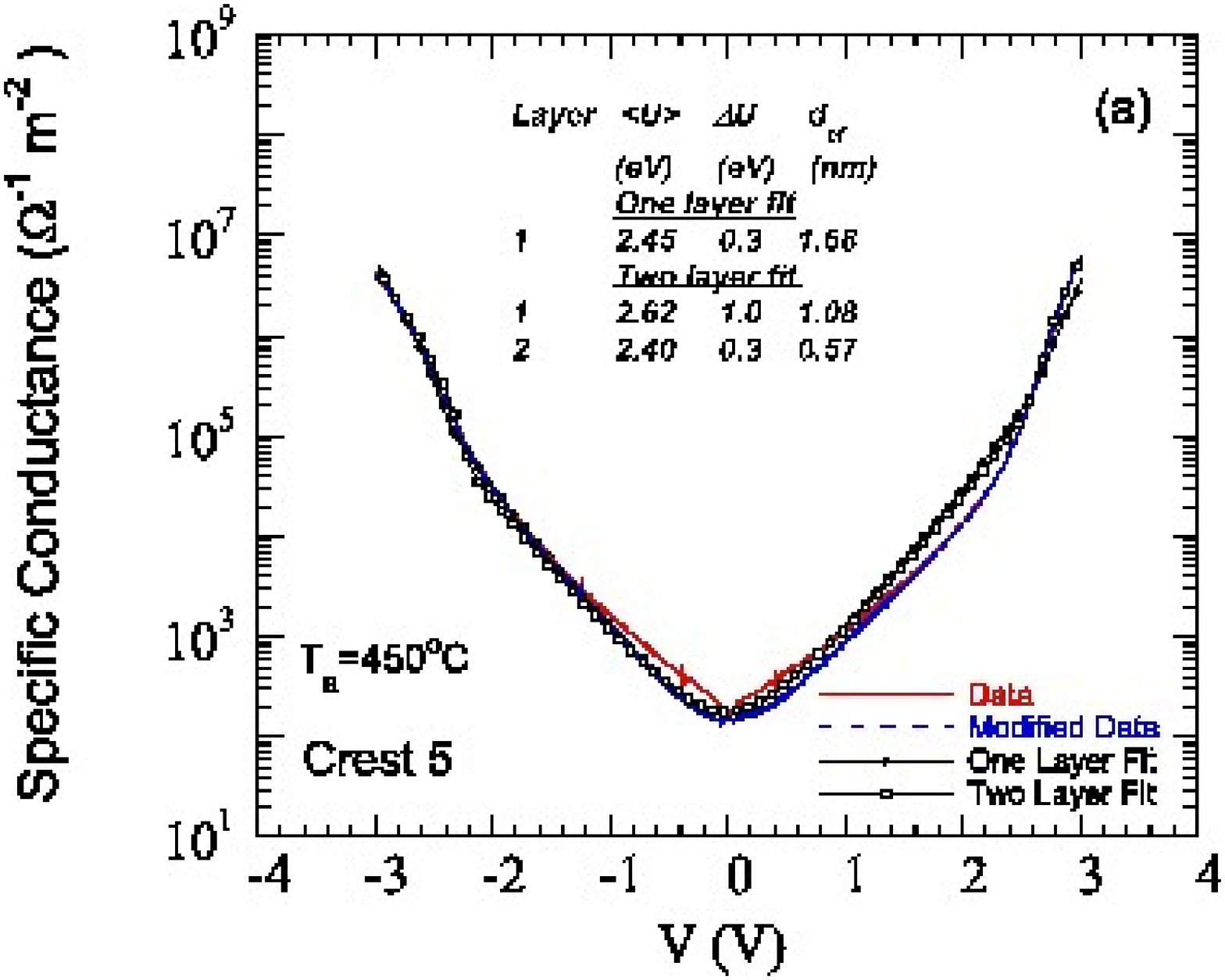}
\includegraphics[width=6.5cm,angle=270]{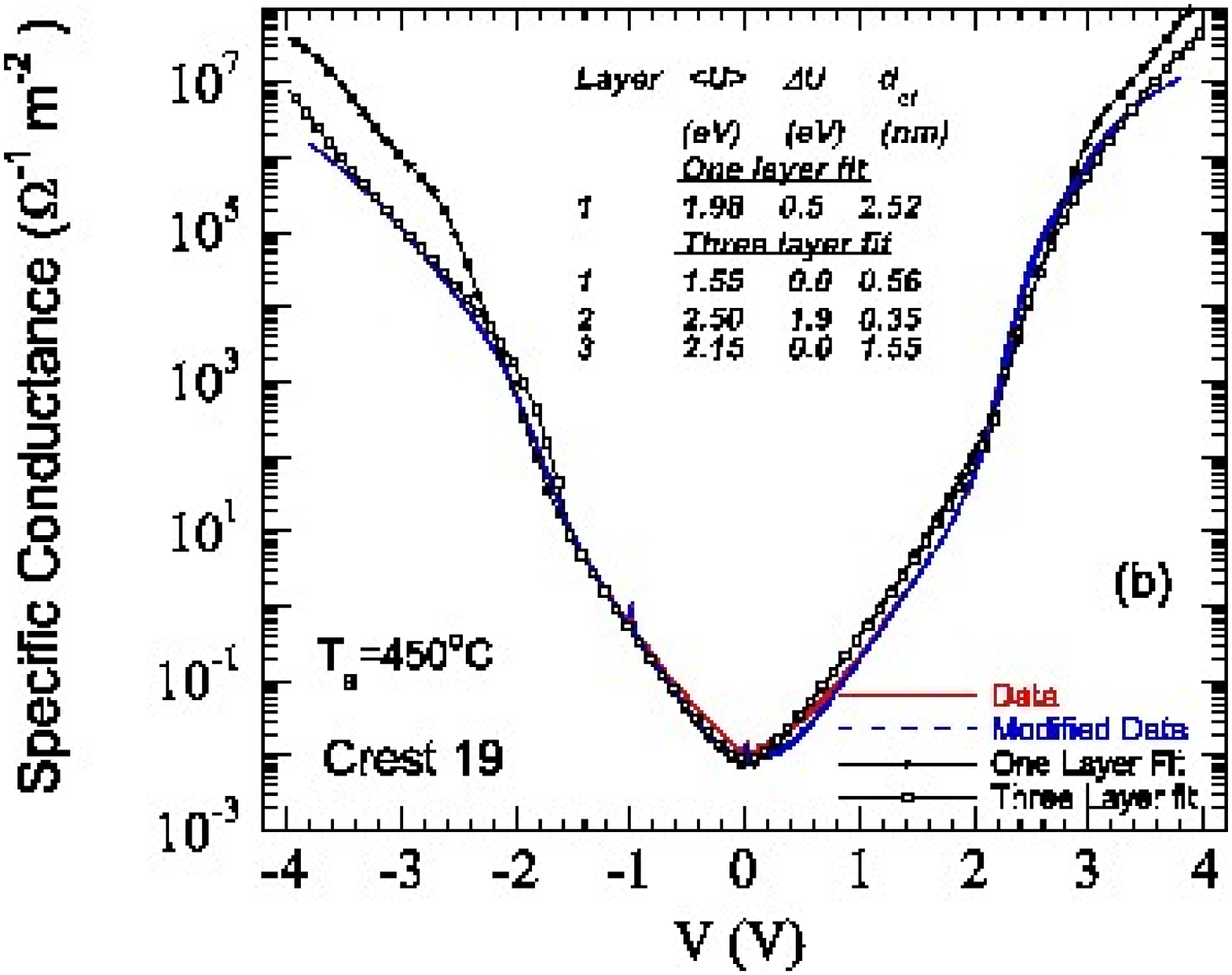}
\caption{Fitting of the specific dynamic conductance $g(V)\equiv A^{-1}dI(V)/dV$ of post-annealed (450$^{\circ}$C) junctions with (a) thermally-grown and (b) plasma-grown barriers with microscopic theory of direct tunneling.  Solid lines show raw data, dashed lines - the data corrected for trap-assisted tunneling (see the text), curves with solid points show the best fits with one-layer (trapezoidal) model, while those with open points for more complex potential profiles.  The fitting parameters (the average barrier height $\langle U \rangle$, asymmetry $\Delta U$ , and effective thickness $d_{ef} = (m/m_0)^{1/2}d)$ are listed inside for each layer, from the base electrode up). }
\end{figure}

In order to estimate $d$ (and hence $m$), we have used measurements of specific capacitance $C_0$ of the annealed junctions (at $T_a$=450$^{\circ}$C). The specific capacitance has turned out to be close to 2.8$\pm$0.7 mF/cm$^2$ for Crest 5 and 2.3$\pm$0.5 mF/cm$^2$ for Crest 19.  Assuming that the dielectric constant of the aluminum oxides is within the range 9$\pm$1 (cited in most publications), the capacitance values imply that the physical thickness of oxides is 2.85$\pm$0.25 nm for the thermal growth and 3.45$\pm$0.25 nm for the plasma oxidation.  These estimates have been confirmed using high-resolution transmission electron microscopy (HRTEM).  For example, Fig. 3 shows two images of a representative Crest 5 junction annealed at 450$^{\circ}$C.  The picture quality is affected by the fact that the base Nb electrode is relatively thick and polycrystalline, so its surface is uneven at a-few-nm scale.  Nevertheless, the images reveal an amorphous AlO$_x$ layer with a thickness of  $\sim$3 nm, i.e. reasonably close to that extracted from capacitance measurements.

Using the effective thickness determined by our fitting procedure (see Fig. 5) we estimate the effective mass (0.35$\pm$0.20)$m_0$ for the thermally-grown oxide and (0.50$\pm$0.15)$m_0$ for the plasma-grown oxide.  These values are in a reasonable agreement with the theoretical result 0.4m0 of Ref. [56], but substantially somewhat lower than the value $\sim 1.0 m_0$, which may be deduced from the experimental results of Ref. [58], assuming that the average barrier height for those films (thermally-grown with UV stimulation and then annealed at 250$^{\circ}$C) is the same as for our thermally-grown layers annealed at the similar temperature.  (Probably, the reason of the discrepancy is that the above assumption is incorrect, i.e. that the UV stimulation increases the barrier height substantially.)

We have applied the fitting procedure described above to extract the average barrier height for both as-grown and post-annealed aluminum oxides. We found that the average barrier height $\langle U \rangle$,  of the thermally-grown oxide increases rapidly at annealing temperatures above 300$^{\circ}$C: from an initial value of $\sim$1.8 eV\cite{22} to $\sim$2.45 eV at 450$^{\circ}$C, and remains close to this value for all the higher annealing temperatures we have explored (up to $\sim$550$^{\circ}$C).  On the other hand, the average barrier height of the plasma-grown oxide remains practically unchanged at around 2 eV.  Semi-quantitatively, this is directly visible from the high-V experimental data shown in Fig. 1, since the barrier height (expressed in electron-Volts) is always close\cite{20} to the voltage of the maximum positive curvature of semi-logarithmic plots $lnI$ vs. $V$, corresponding to the crossover between tunneling through the barrier as a whole at lower voltages, and the Fowler-Nordheim tunneling through its unsuppressed part at higher $V$.

Thus the average barrier height for thermally-grown, post-annealed aluminum oxide layers is substantially (by ~25\%) higher than that in the plasma-grown layers.  This fact offers the possibility of using the oxides in layered (e.g., "crested") barriers for advanced floating-gate memories and other applications.\cite{1}  Figure 6 shows the tunnel current density $J\equiv I/A$ and the corresponding time scale $\tau$ of floating gate recharging calculated for two promising layer combinations:
	(i) thermal oxide similar to annealed Crest 5, plus plasma oxide similar to Crest 19, and
	(ii) 1.25-nm SiO$_2$ layer, plus AlO$_x$ layer similar to Crest 19.

The plots show that the all-aluminum layered barrier of type (i) may sustain the 10-year retention time (standard for nonvolatile memories) at voltages below 1.5 V, while the voltage increase to $\sim$4 V (i.e., by a factor less than 3, enabling a simple NOR structure of memory blocks5) would cause the gate recharging in $\sim$10 $\mu$s.  Such write/erase time is still too long for RAM applications.  Notice, however, that the voltage applied to each of the layers would be below 2.2 V, ensuring high endurance: charge-to-breakdown well above $10^5$ C/cm$^2$, corresponding to more than $10^6$ re-write cycles.  This option may be attractive for low-voltage flash memories, especially because there are good prospects of increasing the barrier endurance even further by using higher post-annealing temperatures\cite{58} and/or Zr alloying of the barriers.\cite{59}

The results for option (ii), i.e., SiO$_2$/AlO$_x$ barriers, are even more interesting.  At $V$ = 3.2 V (or higher) such a barrier would allow the floating gate to recharge in less than 1 nanosecond, with voltage about 1.6 V across each layer.  For electric fields that are so low, we could not even measure the charge-to-breakdown experimentally, but a simple extrapolation of the high-V data gives an estimate of $\sim$10$^{15}$C/cm$^2$, corresponding to $\sim$10$^{11}$ re-writing cycles, which are sufficient for RAM applications.  The drawback of these barriers would be a relatively short retention time ($\sim$100 s at 1.5 V).  Too short for nonvolatile memories, this time is still sufficiently long for DRAM-like memories with periodic refresh.

\begin{figure}[tbh]
\includegraphics[width=6.5cm,angle=270]{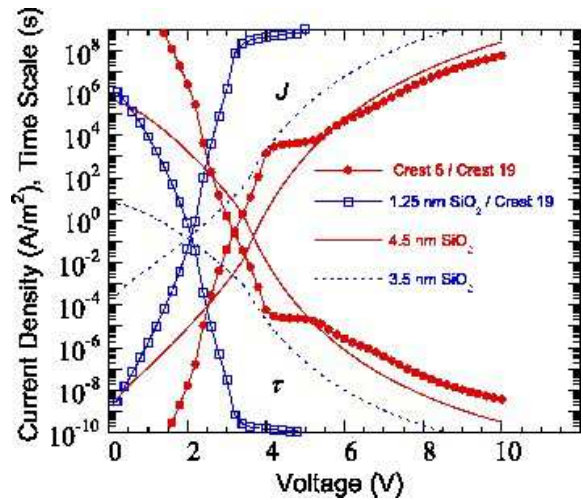}
\caption{Tunnel current density $J=I/A$(increasing with the applied voltage $V$) and the recharging time constant $\tau \equiv C_0V/J(V)$ (decreasing with voltage) for two layered tunnel barriers and two uniform SiO$_2$ barriers, calculated using the aluminum oxide parameters shown in the inset of Fig. 5 and for silicon dioxide parameters taken from Ref. 48 ($U$=3.34 eV, $m/m_0$ = 0.35).  The used dielectric constant values are 10 and 3.9, respectively.}
\end{figure}

These estimates should be, of course, looked upon with caution, since the calculations shown in Fig. 6 imply that the two layers, which had been grown and measured separately in our experiments, may be combined without a substantial change of their properties.  It is more probable that the sequential deposition of the layers will cause at least a moderate change of their parameters and, hence, a deviation from these predictions.  Note, however, that these changes may be either detrimental or beneficial for the crested barrier properties.  Moreover, some barrier parameters (e.g., thickness of the plasma-grown layer) can be easily changed to compensate for undesirable barrier alterations and to improve the crested barriers performance even further. 

To summarize, we have found experimental evidence that electron transport through thermally- or plasma-grown, post-annealed ultrathin aluminum oxide layers is dominated by direct tunneling in electric fields up to $\sim$10 MV/cm.  The effective height of the corresponding tunneling barriers, within the annealing temperature range from 300$^{\circ}$C to 550$^{\circ}$C, is substantially dependent on whether the layer has been grown by thermal or plasma oxidation.  This fact offers hope for the implementation of layered all-AlO$_x$ and SiO$_2$/AlO$_x$ barriers for advanced floating-gate memories.  Our plans are to explore such barriers experimentally in near future.

This work was supported in part by AFOSR.  The authors are grateful to Yu. A. Polyakov and X. Wang for technical assistance.  Useful discussions with A. Bratkovsky, J. Cosgrove, M. Gribelyuk, M. Gurvitch, E. P. Gusev, and T.-P. Ma are gratefully acknowledged.


\begin{thebibliography}{99}
\bibitem{43} While the run-to-run reproducibility of annealing results was very good, and we could get very similar results for annealing in both Ar and N$_2$, the gas choice has unfortunately resulted in a substantial shift: in order to get the same annealing effect in argon, the perceived temperature should be approximately 100$^{\circ}$C higher than for nitrogen, apparently because of a different gas pressure.  This is why the temperatures cited below (that are the averages between the readings for Ar and N$_2$) should be considered as uncertain by approximately $\pm$50$^{\circ}$.
\bibitem{55} In fact, the expression $d_{ef} = (m/m_0)^{1/2}d$ is exact only within the WKB approximation. We have checked that the microscopic theory gives results that may be approximated with the formula $d_{ef} = (m/m_0)^{\beta}$ with the values of $\beta$ very close to 0.5 ($\sim$0.506 for the parameters inferred for sample Crest 5 and $\sim$0.502 for Crest 19).
\end{thebibliography}
\end{document}